\documentclass[useAMS,usenatbib]{mn2e}

\usepackage{graphicx,times}
\usepackage{colortbl}
\usepackage{soul,color}

\def\aj{{AJ}}

\def\msun{M_\odot}
\def\ltorder{\mathrel{\raise.3ex\hbox{$<$}\mkern-14mu
             \lower0.6ex\hbox{$\sim$}}}

\def\r1{$r_1$}
\title[Monte Carlo Simulations of Star Clusters - {}{{VII}}. The globular
  cluster 47 Tuc]{Monte Carlo Simulations of Star Clusters -
  {}{{VII}}.  The globular cluster 47 Tuc}
\author[M. Giersz and D.C. Heggie]{Mirek Giersz$^{1}$\thanks{E-mail:
mig@camk.edu.pl (MG); d.c.heggie@ed.ac.uk (DCH)} and Douglas
  C. Heggie$^{2}$
\\
$^{1}$Nicolaus Copernicus Astronomical Centre, Polish Academy of Sciences, ul. Bartycka 18, 00-716 Warsaw, Poland\\
$^2$School of Mathematics and Maxwell Institute for Mathematical
 Sciences, University of Edinburgh, King's Buildings, Edinburgh EH9
 3JZ, UK}
\begin{document}

\date{Accepted \ldots. Received \ldots; in original form \ldots}

\pagerange{\pageref{firstpage}--\pageref{lastpage}} \pubyear{2002}

\maketitle

\label{firstpage}

\begin{abstract} 
We describe Monte Carlo models for the dynamical evolution of the
massive globular cluster 47 Tuc (NGC 104).  The code includes
treatments of two-body relaxation, most kinds of three- and four-body
interactions involving primordial binaries and those formed
dynamically, the Galactic tide, and the internal evolution of both
single and binary stars.  We arrive at a set of initial parameters for
the cluster which, after 12Gyr of evolution, gives a model with a
fairly satisfactory match to surface brightness and density profiles, 
the velocity dispersion profile, the
luminosity function in two fields, and the acceleration of pulsars.
Our models appear to require a relatively steep initial mass function
for stars above about turnoff, { with an index of about 2.8 (where
the Salpeter mass function has an index of 2.35), and a relatively flat initial mass
function (index about 0.4)} for the lower main sequence.  According to the model, the current
mass is estimated at 0.9 million solar masses, of which about 34\%
consists of remnants.  We find that primordial binaries are gradually
taking over from mass loss by stellar evolution as the main dynamical
driver of the core.  Despite the high concentration of the cluster,
core collapse will take at least another 20Gyr.

\end{abstract}


\section{Introduction}

This is { one of } a series of papers in which we  attempt to
construct dynamical evolutionary models of old star clusters in the
Milky Way.  The first in the series concerned the massive, southern
cluster $\omega$ Centauri \citep{GH2003}.  That study was a pilot
project which used a more primitive version of the code than is
available now, especially with regard to stellar evolution.
Nevertheless it yielded quantitative theoretical results on mass
segregation, which is usually neglected in studies of this cluster
because of its very long relaxation time.  Our updated code, more or
less in its present form, was then tested in application to the old
open cluster M67 \citep{GHH2008}, by comparison with existing $N$-body
modelling.  Then we returned to the globular clusters, for which
$N$-body modelling is either impossible or very limited.  Our model of
the nearby cluster M4 \citep{HG2008a} showed that this object appears
to have passed through core collapse, despite its uncollapsed
(King-like) surface brightness profile.  Turning next to the rather
similar object NGC6397 \citep{GH2009}, which does have a profile of
the type thought to be typical of a post-collapse cluster, we found
that it should be in a similar evolutionary phase to M4, and concluded
that the difference between the two clusters was best explained as a
result of fluctuations.  Detailed $N$-body modelling \citep{HG2009}
confirmed that a post-collapse cluster like NGC6397 should exhibit
oscillations, on a time scale of order $10^8$yr, as a result of which
it could sometimes look like M4, and sometimes like NGC6397.  In
summary, each time we have looked at a specific globular cluster with
our technique, we have discovered something slightly surprising{:
  mass segregation in $\omega$ Cen, the postcollapse status of M4, and
  core oscillations in NGC6397}.

In the present paper we turn our attention to the other famous, massive,
southern globular cluster -- 47 Tucanae (NGC104).  Like M4 and NGC6397
there is a wealth of observations, even though it is considerably more
distant.  It is also much richer, reducing some statistical errors in
such data as the surface brightness distribution.  It was selected for
detailed study during discussion at the programme ``Formation and Evolution
of Globular Clusters'', held at the Kavli Institute for Theoretical
Physics, Santa Barbara, California, USA, in January 2009, just as M4
had been selected at the MODEST-5 meeting (``Modelling Dense Stellar
Systems'') at McMaster University, Hamilton, Canada, in August 2004.
From the point of view of dynamical evolution, it differs from most of
our
earlier targets because of its long relaxation time (which is a
consequence of its much greater mass).  This makes it dynamically
younger, raising the expectation that it should be easier to model.
But the long relaxation time should also imply that its present state is more
directly influenced by its initial conditions.

This paper first reviews both the observational data on 47 Tuc and
earlier work on its dynamics (Sec.2).  { Then in Sec. 3 we discuss our
techniques, in terms of  the simulation of dynamical evolution, {
  the calculation of mock observational data} and
our search for satisfactory initial conditions.  In the following
section, Sec.4, we present our optimal model, in the sense that, among
the models we have studied, it is representative of those that provide
the most satisfactory fit to a wide selection of the available
observational data.  Finally (Sec.5) we consider such issues as the
plausibility of our initial conditions, how these  relate to current
ideas on the earlier stages of cluster formation and evolution, and
the mechanisms which dominate the dynamical evolution of the cluster
at the present day.  Section 6 sums up.}

\section{A review of 47 Tuc}

\subsection{Dynamical models}

As with all globular star clusters, dynamical models { of 47 Tuc} fall into two
classes \citep{HG2008b}: static models, such as King models
\citep{Ki1966}, and dynamical evolutionary models, which are the focus
of our own contribution.  (We exclude here the practice of fitting a
template to such data as the surface brightness profile.  Though often
referred to as King models, and useful as they are for the measurement
of such parameters as the core radius \citep[e.g.][]{Ca1993}, they have
no real dynamical basis.)  In this subsection we survey existing
models of these two kinds, though we often have to refer to various
kinds of observational data on which we expand in later subsections.

\citet{II1976} fitted a single-mass King model to a composite surface
brightness profile constructed from photometric data and star counts.
Simple though it is, such a model was also used more recently by
\citet{Fr2001} for a study of pulsar accelerations; and by
\citet{Mc2006} for the interpretation of the velocity dispersion
profile, derived from proper motions and radial velocities. 
Several investigators have constructed multi-mass generalisations of
the King model for 47 Tuc \citep{DF1985,MM1986,Ma2004}.
\citet{Me1988,Me1989} and \citet{Re1996} added anisotropy, in the style of
\citet{Mi1963}.

A number of other, rather different static models have been constructed for
47 Tuc.  It was pointed out by \citet{SP2001} that a better fit to the
surface brightness profile could be obtained by construction of a
model with a slightly different choice of distribution function from
that of King.  The rotation of 47 Tuc has been the subject of
considerable study, but almost all dynamical models ignore it.  One
exception is the pair of models constructed by \citet{Da1986}.
Finally in this section we mention the non-parametric analysis of
\citet{Ge1995}, which {\sl derives} the distribution function from the
observational data.

More in line with the aims and methods of this paper are the dynamical
evolutionary models of \cite{Mu1998} and \citet{Be2003}.   Though
few details are available in these abstracts, the
technique is a multi-mass Fokker-Planck code, with a discretised, piecewise
power-law mass function and heavy remnants.   The models are able
to account for the stellar mass function, star-counts, the velocity
dispersion profile, pulsar accelerations, and the radial distribution
of neutron stars.  { The models fit} better than a King-Michie model.

\subsection{Observational Data}\label{sec:od}

{ Our purpose in this subsection is to review something of the wealth of
observational data which exists for 47 Tuc.  Not all of this data is
compatible, and one of our aims is to explain where and why we have
selected one dataset over another. }



\subsubsection{Surface brightness profile and star counts}\label{sec:sbp-observational}

In several of the foregoing investigations the spatial distribution of
matter in 47 Tuc has been specified in terms of a surface
brightness profile, and two major compilations exist
\citep{Me1988,Tr1995}.  The later of these is largely a superset of
the former, though \citet{Me1988} includes star count data in J from
\citet{DC1982} which are excluded from the later compilation.  The two
compilations agree quite well; \citet{Tr1995} also provide an analytic
fit to their data, and the rms difference between this and the data in
\citet{Me1988} is 0.28 mag.  A large fraction of this is contributed
by star counts at large radius, including the aforementioned J data.

Despite this apparently satisfactory state of affairs, some caution
must be exercised in the construction of a composite profile from such
disparate data.  Da Costa himself (his Fig.7) shows that the slope of
the density profile from different plates (differing essentially in
the range of masses of the counted stars) is different, in a manner
consistent with mass segregation.  A further complication is the
colour gradient exhibited by 47 Tuc \citep{CF1979}, which has been
traced to a radial variation in the fraction of light contributed by
{\sl bright} giants \citep{Fr1985}.  One implication of these results
is that it is not strictly possible to reconstruct the surface
brightness from star counts by a constant shift, as is done in these
compilations.  Furthermore, beyond about $5'$ the only genuine
photometric data in these compilations are the drift scan measures of
\citet{GB1956}, and an inspection of their Fig.1 does not inspire much
confidence in the results at these radii.
With this consideration in mind, in our modelling we
have not relied solely on the surface brightness profile of
\citet{Tr1995}, but have also compared our models with counts from two
sources,  (i) the
two deepest V data in the compilation of \citet{Tr1995}, i.e. plates
3407 and 3754 from \citet{DC1982}, and (ii) the  surface density profile
(based on star counts for stars above turnoff mass in the innermost
$82.5''$)  published by \citet{Mc2006}.

{ One significant detail about our handling of the data in
\citet{Mc2006} concerns the passband.  \citet{Mc2006} give surface
densities for stars above about turnoff; more precisely, for stars
with $m_{\rm F475W} < 17.8$, corresponding roughly to $V = 17.65$
  \citep[Sec.4.1]{Mc2006}.  These are empirical values, whereas our
model creates $V$ magnitudes based on model atmospheres.  We have
checked several synthetic databases, and find that $m_{\rm F475W} - V$
is considerably larger.  For example for Padova isochrones
(http://stev.oapd.inaf.it/cgi-bin/cmd) it is 0.284, for the bluest
stars around turnoff at the appropriate age and metallicity, and for
the appropriate Dartmouth models it is still larger.  In
Sec. \ref{sec:sbp} we present results based on both the empirical and
synthetic values.}






















\subsubsection{The velocity dispersion profile}\label{sec:observed-vdp}

For velocities we initially relied on the
line-of-sight velocity dispersion profile of \citet{Me1988}, which was
based on a catalogue of 169 radial velocities of giants.
Nevertheless much larger data sets have been acquired since, and we
begin this section by reviewing more recent work, with a view to
determining if the older data is adequate for our purpose.

Considerable interest has centred on the {\sl subsequent} discovery \citep{Me1991} of 
two high-velocity stars with
line-of-sight speeds (relative to the mean of the cluster)
approximately 4 times larger than the central line-of-sight velocity
dispersion.   { It can be argued }
that there is nothing particularly abnormal about these
\citep{Pe1994,Mc2006}.   They do, however, have the effect of increasing the central
line-of-site velocity dispersion from 9.4$\pm1.0$km/s \citep{Me1991} to
11.6km/s.  Therefore it would not be surprising if the central
velocity dispersion of a successful model exceeds that of the older data.

  \begin{figure}
{\includegraphics[height=12cm,angle=0,width=9cm]{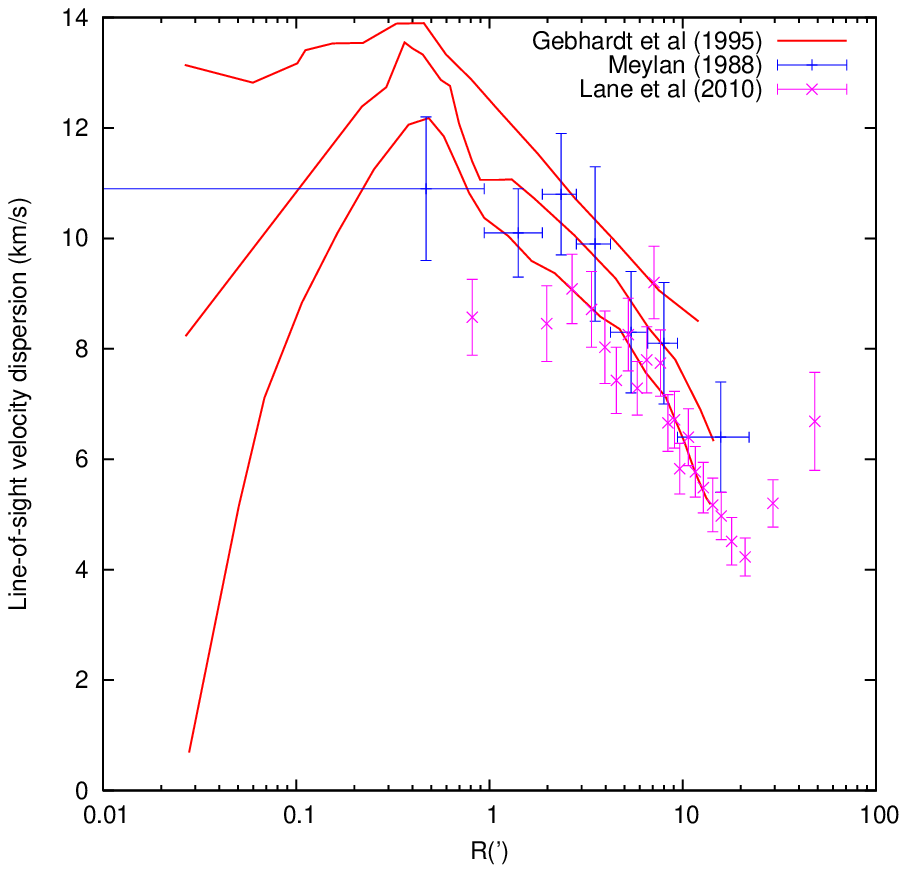}}
    \caption{Line-of-sight velocity dispersion profiles of
    \citet{Me1988}, \citet{Ge1995} and \citet{LKLISBS2010}.  For the
    first of these the vertical error bars
    are as in the original publication, and the horizontal error bars
    give the range of projected radii in which the stars fall.  For
    \citet{Ge1995} we have copied curves in their figure which give estimates
    of the velocity dispersion profile and its 90\% confidence
    interval.  {For the most recent data set \citep{LKLISBS2010} we have copied the points and
    vertical error bars from their Fig 11.}}
\label{fig:rv}
  \end{figure}

Radial velocities for 548 stars in the central region of 47 Tuc have
been secured by \citet{Ge1995}, and the inferred velocity dispersion
profile is plotted 
along with the older data in Fig.\ref{fig:rv}.  This figure confirms
the result that the central velocity dispersion may be considerably higher
than in \citet{Me1988}.   

{Recently 
{\citet{LKLISBS2010} obtained
a velocity dispersion profile based on the largest spectroscopic sample ever
observed for 47 Tuc \citep[see also][]{Ki2007,Sz2007}, and we have
included the results in the figure. Their results show that for the outer parts of the
cluster the velocity dispersion appears to be rising with radius,
while the central value ({obtained by fitting a Plummer model}) is
only about 9.6 km/s. 
{ Note, however, that one of the criteria for membership of 47 Tuc
  { in their study} was
that the radial velocity must lie within a certain range of width
about 45 km/s (their Fig.1).  This would have excluded the two
high-velocity stars discovered by  \citep{Me1991}, and probably
suppresses the velocity dispersion at small radii.  Twenty-five stars
which, according to the criteria of membership in \citet{LKLISBS2010},
are members of 47 Tuc, actually lie outside the conventional tidal
radius.  Several of these extratidal stars lie in sparsely populated
parts of the colour-magnitude diagram (their Fig.3), { which
  suggests that they may not be members at all.}   It is
possible that the membership criteria are insufficiently strict at
large radii.}}}

{
On account of such arguments,  for the work reported here we have relied mainly on the data
of Gebhardt et al, and so we now discuss it a little further.}
{ The sharp drop in the Gebhardt et al data just inside 1 arcmin is
interesting.  \citet{Ge1995} regarded it as a sharp rise (as the
radius decreases), and suggested tentatively that it may indicate a
central population of dark objects.  The analysis of \citet{GF1995},
however, implies that the M/L ratio inside this feature is little
higher than at 5 arcmin.  In fact it implied that there was a zone of
very {\sl low} M/L ratio around 1 arcmin (presumably because the drop
in velocity dispersion in this vicinity is almost Keplerian).  { Furthermore}
the 90\% confidence limits on this result require an increase in M/L
with {\sl increasing} radius between about 1 and 4 arcmin.  The lesson
we draw from this discussion is that the features in the velocity
dispersion profile may have little significance, and indeed the
confidence limits themselves are subject to some doubt.}

In our selection of models (Sec.\ref{sec:models}) we have not used the
wealth of data on proper motions \citep{Re1996,KA2001,Mc2006}, 
the bulk of which is confined to the innermost $100''$.
Nevertheless we shall compare some aspects of these data, such as
anisotropy (Sec.\ref{sec:va}), with one of our best  models.

The rotation of the cluster is evident in both proper motions and
radial velocities, and has been studied in many of the above papers.
{ Unfortunately, however, our Monte Carlo technique is restricted
to the case of spherical symmetry and no rotation.  We return to this
issue in Sec.\ref{sec:vdp}.}

\subsubsection{Luminosity and mass functions}\label{sec:lf}

Ground-based data on the mass function \citep{He1987} provided an
important constraint for the fitting of King-like models
\citep{Me1989}, and we shall see that luminosity functions continue to
serve such a role in our work.  { Indeed the use of the luminosity
  function is preferable, as it places the comparison closer to the
  observational domain, and simplifies problems connected with the
  mass-luminosity relationship.}  We have, however, mainly used more
recent data:  the 
luminosity functions derived from HST observations  by \citet{AK1996} at
1 and 14 core radii, where the core radius was taken to be $23''$.
The outer field goes down to magnitudes corresponding to stellar
masses of about $0.1\msun$.  Other information on the luminosity
function, though not used here, can be found in \citet{dP1995,Sa1996}
and \citet{Ho2000}.

{ We have retained the older data of \citet{He1987} for one discussion
in Sec. \ref{sec:lf-a}, where we consider their ``composite''
luminosity function.  As the authors themselves state, they did not
attempt to correct for mass segregation when combining data secured at
different radii, and it should not necessarily be regarded as a global
luminosity function.  Our interest in it is focused on the stars above
turnoff, however.}


\subsubsection{Binary fraction}\label{sec:pbs}

{ The binary fraction is } very relevant for a variety of exotic objects, including blue
stragglers and millisecond pulsars.  { Nevertheless it}
seems to be observationally rather poorly constrained in most globular
clusters, and 47 Tuc is no exception.  The issue is complicated by
mass segregation and the observational selection effects of the
various techniques for detecting binaries.  

From inspections of colour-magnitude diagrams, \citet{dP1995} and
\citet{Sa1996} estimated that the fraction of ``equal-mass'' binaries
in fields $4.6'$ and $5'$ from the cluster centre was at least 5\%.  Using
the same basic approach, at the same distance of $4.6'$, \citet{An1997}
found a fraction of at most 2\% with mass ratio $q> 0.5$, in the
magnitude range from V = 20.5 to 23.5.  Again using broadly the same
technique, Milone (pers.comm.) finds the fraction of binaries with $q>0.5$ to
be $0.015\pm0.008$, and the fraction of all binaries to be
$0.026\pm0.015$.  His measures refer to a region between 0.95 and 2.4
arcmin from the centre, and he points out that they take no account of
the recently discovered intrinsic broadening of the main sequence
\citep{An2009}. 
In the central regions
($<90''$) \citet{Al2001} estimated a binary fraction of
about 14\%, based on observations of eclipsing binaries and some
modelling.  The stars observed range in $V$ from about 16 to about 23,
and it is known from modelling (e.g. \citealt{HG2008a}) that mass
segregation is particularly pronounced among bright objects.
\citet{Kn2008} estimated a similar binary fraction among white dwarfs
in the core.

{ Even though these estimates refer to somewhat different populations of
  binaries, and in some cases at different locations, the binary
  fraction seems to be the least well constrained of the observational
  quantities we have discussed.}



\subsubsection{Pulsars}\label{sec:pulsars}

It is estimated \citep{He2005} that 47 Tuc contains about 300 neutron
stars.  Among these are 23 observed millisecond pulsars
\citep{Fr2003}\footnote{Updated at
http://www2.naic.edu/$\sim$pfreire/47Tuc/}.  Since our modelling focuses on
dynamical issues, the interest of these objects is two-fold.  First,
some previous models of 47 Tuc have invoked such heavy remnants as
being necessary to account for the central velocity dispersion
\citep[e.g.][]{MM1986}.  Second, well-observed pulsars act as probes of the
gravitational field \citep{Ph1992}, { in a way that we now summarise.}

Sixteen of the millisecond pulsars have timing solutions which
include measurements of the logarithmic period derivative $\dot P/P$,
where $P$ is the { pulse} period.  One contribution to this is intrinsic, {
  caused by the spin-down of the pulsar}, but the size of
this contribution (which is positive) is estimated to be of order
$(\dot P/P)_{int}\sim0.5\times10^{-17}$/sec \citep{Fr2001}.  In
addition there is a Doppler contribution from the relative  acceleration of
the pulsar and the observer, and the main contribution is generally
that due to the matter in the cluster itself.  { For a spherically
  symmetric cluster the magnitude of this term is} ${GM(r)}\cos\theta/{(cr^2)},$
where $M(r)$ is the cluster mass within the radius $r$ of the
pulsar, $G,c$ are the usual physical constants, and $\theta$ is the
angle between the line of sight to the pulsar and the outward radial
direction at its location in the cluster.  Thus the spin period of a
pulsar located on the far side of the cluster appears to be
decreasing, unless the contribution from intrinsic spin-down is too
large.  Of course only the projected radius $r_p = r\sin\theta$ is known 
with any accuracy.  For a given cluster model and given $r_p$,
however, the extreme values of the { Doppler contribution}
can be calculated.  Then, all observed values of
$\dot P/P$ should lie between these extrema, except for the small
intrinsic contribution.  This is illustrated for one of our models in
Fig.\ref{fig:pulsars-a}. 


\section{The simulations}\label{sec:models}

\subsection{Model description}\label{sec:spec}

As in previous papers in this series, the code we use is an updated
version of the Monte Carlo code developed in
\citet{Gi1998,Gi2001,Gi2006}.  { It includes} synthetic stellar
evolution of single and binary stars using prescriptions described by
\citet{Hu2000} and \citet{Hu2002}.  The only relevant { updates} to report here
are the inclusion of (i) exchange interactions (through the use of
cross sections), and (ii) natal kicks for both neutron stars and black
holes (which had been confined to single neutron stars in previous
versions of the code).  

\begin{table*}
  \begin{minipage}{18cm}

  \begin{center}
\caption{Choice of initial conditions etc}
    \begin{tabular}{lllllll}
Property& Distribution adopted& Parameter & Meaning &Range (small-&
Range (full-&Notes\\
&& & &scale models)&
scale models)&\\
\hline
Number of stars&&$N$&&$a$&$1.8-2.2\times10^6b$\\
Structure&\citet{Ki1966}&$W_0$& Central potential&$4-11$&$7-10$&$c$\\
Tide&Giersz et al (2008)&$r_t$&Tidal radius&$a$&$72-116$ pc&See text
Sec.\ref{sec:spec}(iv) and $d$\\
Radius parameter&&$r_t/r_h$&&$25-500$&$21-60b$&Determines $r_h$ (given $r_t$)\\
Mass function&Two-part power law& $m_2$&Maximum
mass&$2-150\msun$&$2-150\msun$&\citet{Kr2008} but with \\
(single stars)&& $m_b$&Break mass&$0.86-0.9\msun$&$0.75-1\msun$&different parameter values\\
&& $m_1$&Minimum mass&$0.1\msun$&$0.08\msun$\\
&&$\alpha_2$&Upper index&$2.3-5.3$&$2.3-4.5$&Salpeter is 2.35\\
&&$\alpha_1$&Lower index&$0.4-1.3$&$0.4-1.3$&\\
Binaries&\citet{Kr1995}&$f_b$&Binary fraction&$0-0.20$&$0-0.06$&See text Sec.\ref{sec:spec}(ii)\\
Mass segregation&None&&&&&But see the end of Sec.\ref{sec:vdp}\\
Natal kicks&Gaussian&$\sigma$&1-dimensional dispersion&$190$ km/s&$130-190$ km/s&neutron stars
and black holes\\
Age&&&&11 Gyr&$11-12$ Gyr&\\
Metallicity&&$Z$&&0.0035&$0.003-0.0035$&\\
\hline
    \end{tabular}\label{tab:ic}
  \end{center}
Notes: 
$a$ See text Sec.\ref{sec:finding-a-model}.
$b$ Larger values were tried for truncated polytropic models.
$c$ Also \citet{Wo1954} and polytropic models (Sec.\ref{sec:vdp}).
$d$ A wider range was tried for truncated polytropic models.
  \end{minipage}
\end{table*}

For initial conditions we adopt the  assumptions { given in Table
\ref{tab:ic}.  The table also includes an indication of the range of
numerical values which were explored in the search for a model
{ which is} described in Sec.\ref{sec:finding-a-model}.  
Some of our choices  require a little elaboration here.}
\begin{enumerate}
\item  { The initial
mass function (a two-part
power law) depends on five parameters, but only four were adjusted in
the search for a model.  For the low-resolution initial search
(Sec.\ref{sec:finding-a-model}) the minimum mass was held at $m_1 =
0.1\msun$, as in previous papers in this series.  It was kept at $m_1
= 0.08\msun$ for the full-resolution models discussed in
Sec.\ref{sec:modela}.


\item As in our work on M4 \citep{HG2008a}, the binary parameters follow the prescription of \citet{Kr1995},
  except that the initial component masses are derived from the same
  initial mass function as the single stars in our model; its
  parameters normally differ in value from those in the mass function used by Kroupa.  The
  component masses, period and eccentricity are then subject to
  ``eigenevolution'' and ``feeding'' algorithms.  { These} have a physical
  basis and are designed to build in a number of correlations
  consistent with observational data.}


\item  
For the initial
spatial distribution we have adopted King models, but models which (in
general) underfill the initial tidal radius.  Thus there is a
distinction between the tidal radius (determined by the strength of
the tidal field and the mass of the cluster) and the edge radius of
the King model.  { We have also experimented with several other types of
model, as described briefly in Sec.\ref{sec:vdp}.  }

\item
A constant tidal field strength  is used,
as if the cluster were in a circular orbit.  For 47 Tuc \citet{Tu1992}
and \citet{Di1999} estimated an eccentricity about 0.17,
i.e. the Galactocentric distance varies by approximately this fractional
amount on either side of its mean value.  { Therefore} our assumption has some
approximate validity.  

\item
{ Though { it is} not strictly an issue of modelling, for the purposes of
  comparison between model and observations we need to choose values
  for the age and distance of the cluster.   The adopted distance is
  the value given in {}{{Table \ref{tab:data}}}, but for the main model described in
  Sec.\ref{sec:modela} we found it slightly advantageous to choose an
  age at the upper end of the range in the Table, i.e. 12 Gyr (see Sec.\ref{sec:variations}).}

\end{enumerate}

{

\subsection{Generation of observational data}

We have already discussed (Sec.\ref{sec:od}) the main kinds of
observational data which we shall attempt to fit with our model.  
Here we give the recipes used for the generation of the corresponding
data from our model (and some additional data such as the core
radius).  

At any time, the output of the Monte Carlo model
consists of a list of data for each star or binary.  This data
includes the radius $r$ (i.e. distance from the cluster centre),
radial and tangential velocity ($v_r$ and $v_t$), mass, 
$V$-luminosity $L_V$ in solar units, and colour of the
star.  For binaries the same data is available for both components,
along with the semi-major axis.

\subsubsection{Surface brightness}\label{sec:sb}

To construct the surface brightness at a given projected radius $d$,
we replace each star by a uniform transparent spherical shell of 
radius $r$.  Its surface (projected) luminosity density is given by $\Sigma_V =
\displaystyle{\frac{L_V}{2\pi r^2}\frac{r}{\sqrt{r^2-d^2}}}$ if
$r>d$.  The total surface luminosity density is the sum over all
shells of radius $r>d$ (which we also denote by $\Sigma_V$ in what follows), and then the surface brightness is given (in
$V$ magnitudes per square arcsec) by 
\begin{equation}
\mu_V = \displaystyle{V_\odot -
  2.5\log_{10}\left(\Sigma_V\frac{10^2\pi^2}{648000^2}\right) + A_V}, \label{eq:muv}
\end{equation}
where $V_\odot$ is the absolute $V$-magnitude of the sun and $A_V$ is
the visual extinction to the cluster.

This method has a disadvantage that the brightness is infinite along a
line of sight such that $d = r$ for some shell.  
 It would be possible to construct a realisation of the model by
selecting the full position of each star at random on the
corresponding shell, and then determine the surface brightness in a
manner  which is more akin to the observational approach.  We have
verified that this method produces similar results to the method
described in the previous paragraph, and prefer that  method (i.e. eq.(\ref{eq:muv})), as it
has the merit of introducing no additional source of noise.

For the observational core radius we determine the smallest value of
$d$ such that the surface brightness is half that at the centre.  (We
also refer to a dynamical core radius in Sec.\ref{sec:status}.)

\subsubsection{Surface density profile, binary fraction and luminosity function}\label{sec:sdpbflf}

Our calculation of the surface density (i.e. number of stars per unit
area on the sky) uses the same formulae as for surface brightness, as just
described, except that $L_V$ is omitted, and the conversion is to
number per square arcsec and not $V$ magnitudes.  For the (local)  binary
fraction we compute the ratio $n_b/(n_b+n_s)$, where $n_b$ is the
projected density of binaries, and $n_s$ is the projected density of  single
stars.  (Note, however, that the data in Table \ref{tab:modela} give
the 
{\sl global} binary fraction.)
The luminosity
function is calculated in the same way as the surface density, but
separates  stars according to 
the appropriate magnitude bins.

\subsubsection{Line-of-sight profiles of velocity dispersion and
  proper motion}

Consider the point on a shell of radius $r$ at which it is pierced by the line of sight
of projected radius $d<r$.  We assume the transverse velocity of the
star is randomly orientated in the tangent plane to the shell, and
then it is easy to show that the mean square line-of-sight velocity is
given by $\displaystyle{v_r^2\frac{r^2-d^2}{r^2} +
  \frac{1}{2}v_t^2\frac{d^2}{r^2}}$.  For each shell within the line
of sight, this expression is then weighted by the surface density of the
star, i.e. by a factor $r/\sqrt{r^2-d^2}$, and the weighted average
gives the line-of-sight velocity dispersion.

For the line-of-sight velocity dispersion, stars fainter than absolute
magnitude $V=6$ are excluded from the sum.
For proper motions similar formulae are used, with the appropriate
magnitude limit (Fig.\ref{fig:anisotropy}).

\subsubsection{Pulsar acceleration}

For each line of sight, the line-of-sight component of the
gravitational acceleration is computed at the location of each shell,
and the maximum value found.

}

\subsection{Finding a model}\label{sec:finding-a-model}

The starting point of our search for a model was a consideration of
present estimates of the mass and dimensions of the cluster (Table
\ref{tab:data}).  The half-mass relaxation time is long, which
suggests that the cluster has not lost a very large fraction of its
initial mass.  Indeed the formulae given by \citet{BM2003} suggest
that, even with its present mass, the dissolution time is of order
$10^{11}$yr.  This in turn would suggest that most of the mass 
lost so far by 47 Tuc results from stellar evolution, and that its initial mass
was less than twice its present mass.  For much the same reason we
began by assuming that its initial half-mass radius was not very
different from its present value.  Only in the core is the relaxation
time short enough to suggest extensive evolution by relaxation.

\begin{table}
\begin{center}
\caption{Data on 47 Tuc}
\begin{tabular}{lll}
${R_{GC}}^1$& 7.4 kpc\\ 
${R_\oplus}^1$&4.5 kpc\\
${E_{B-V}}^1$& 0.04\\
$[Fe/H]^1$&$-0.76$\\
Total mass $M^2$&$1.1\times10^6M_\odot$\\
Half-mass relaxation time ${t_{rh}}^1$& $3.0\times10^9$ yr\\
Central  relaxation time ${t_{rc}}^1$& $0.9\times10^8$ yr\\
{ Age$^3$}&$11^{+1}_{-1}\times10^9$ yr\\
$r_c^1$&$0.40'$\\
$r_h^1$&$2.79'$\\
$r_t^1$&$42.86'$\\
\end{tabular}
\label{tab:data}
\end{center}
Sources and notes: $^1$ \citet{Ha1996}, updated on\\
http://www.physics.mcmaster.ca/$\sim$harris/mwgc.dat on 19 January 2010; $^2$
\citet{Me1989}; $^3$ \citet{Gr2003}, though their results have been 
simplified slightly
  \end{table}

Guided by these ideas, we began the search for suitable initial
conditions using small-scale models with $N^\ast = 10^4$ or $4\times10^4$
stars.  These were scaled to have the same relaxation time as a
full-scale model with $N$ stars, as described in
\citet{HG2008a}.  Briefly, this is arranged by relating the radial
scales of the two models by 
\begin{equation}
\frac{R_\star}{R} =
\left(\frac{N}{{}{{N}}_\star}\right)^{1/3}\left(\frac{\log(\gamma
  N_\star)}{\log(\gamma N)}\right)^{2/3},\label{eq:scaling}
\end{equation}
where $\gamma = 0.02$.  {(\citet{PZ1999} used a similar trick, but with
$\gamma=1$.)}  For $N^\ast = 10^4$ and $N = 2\times 10^6$
(which is typical of the models that we eventually prefer) the ratio
is $R_\star/R \simeq 3.68$. 

In these small-scale models the age
was taken as 11 Gyr, as in Table \ref{tab:data}, and the minimum mass
was taken to be $0.1\msun$.  These values differ from the values of 12 Gyr
and $0.08\msun$ adopted for the main model in Sec.\ref{sec:modela},
but the resulting differences are negligible for the purpose of the
explorations discussed here. { These models can be scaled to any value
of $N$, which is not an input parameter of the model and is therefore
not given in Table \ref{tab:ic}.  Similarly we do not specify the
range of $r_t$ for these models, as this also depends on the chosen scaling.}

Our first finding was that the canonical values for the initial mass
function did not provide a satisfactory fit to the luminosity
functions which we were trying to match (see Sec.\ref{sec:lf}).  We
found that all our subsequent modelling of the luminosity functions
(at least, below the turnoff) was satisfactory if we chose values
closer to $\alpha_1 =
0.5$ and $m_b = 0.9\msun$, where these parameters are defined in {
  Table \ref{tab:ic}}.   (Nevertheless, slightly different values were eventually adopted for the model presented in
Sec.\ref{sec:modela}; see also Sec.\ref{sec:variations}.)  Thus the low-mass initial mass function is
flatter than in the canonical formulae { of \citet{Kr2008}}.  
Strictly, only the value of $\alpha_1$ is approximately
determined by this comparison; all we can say of $m_b$ is that it is
not much below the present-day turnoff mass, and the value chosen merely
represents approximately the most modest change from the canonical values.

\begin{figure}
{\includegraphics[height=12cm,angle=0,width=9cm]{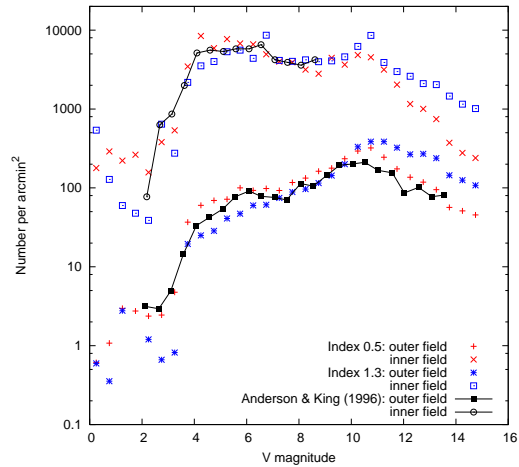}}
    \caption{{Observed luminosity functions at two radii
    \citep[from][]{AK1996} compared with two models.  One is a scaled
    version of the model close to that presented in Sec.\ref{sec:modela}, scaled to
    $N* = 40000$, and the other is an identical model except for the
    power law index of the lower initial mass function, which is the
    ``canonical'' value of 1.3 \citep{Kr2008}.}  The actual initial
    parameters for the first model are given in Table \ref{tab:smallscale}.}  
\label{fig:imf-comparison}
  \end{figure}

Our conclusion about $\alpha_1$ seems quite robust{, and is illustrated in
Fig.\ref{fig:imf-comparison}.  This shows a scaled version of a model
close to the one
 to be presented in Sec.\ref{sec:modela}, and one with identical
parameters (see the caption to the figure and Table \ref{tab:smallscale}) except for the slope of the lower part of the IMF.  Though
a better fit could certainly be obtained by varying more than just one
parameter, the figure does serve to illustrate the effect of a
mismatch in the choice of IMF.}

\bigskip

{ In order to give a flavour of how our search for a model
  proceeded from this point,} we
now explore a coarse grid of the main initial structural parameters, namely
the initial half-mass radius and concentration.  (We adopt the models
of \citet{Ki1966} for the initial structure.)  All the models in this
  particular discussion use a
relatively large initial value of the tidal radius ($r^\ast_t = 500$
pc, where we give the value for the small-scale model.)  It should also be
said that they represent a tiny subset of all the models we explored
  { (see Table \ref{tab:ic})}.

\begin{figure}
{\includegraphics[height=12cm,angle=0,width=9cm]{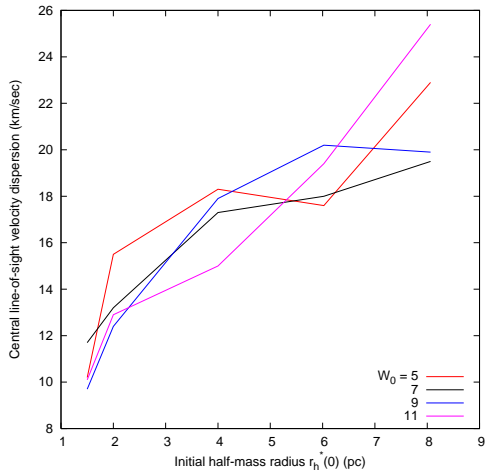}}
    \caption{Line-of-sight velocity dispersion at 1 arcmin for a number of
    scaled models.  Different curves correspond to different initial
    concentrations.  The abscissa is the initial half-mass radius of
    a small-scale model with $N^\ast = 10^4$ stars, and is related to
    that of a full-scale model with $N$ stars by
    eq.(\ref{eq:scaling}).  The models have been scaled to the correct
    central surface brightness for 47 Tuc, and so the value of $N$
    differs from one model to another.}
\label{fig:survey}
  \end{figure}

Some results of this representative mini-survey are
summarised in Fig. \ref{fig:survey}. { The results of each model were
scaled (by the particle number) to fit the central surface brightness
of 47 Tuc}, and then this figure  gives the line-of-sight
velocity dispersion at a projected radius of 1 arcmin.  While these suggest that there
is a range of relatively compact initial conditions ($r_h^\ast$ in the
range from 1.5 to 2 pc) yielding approximately the correct 
velocity dispersion (see Fig.\ref{fig:rv}), further inspection reveals
serious problems.

\begin{figure}
{\includegraphics[height=12cm,angle=0,width=9cm]{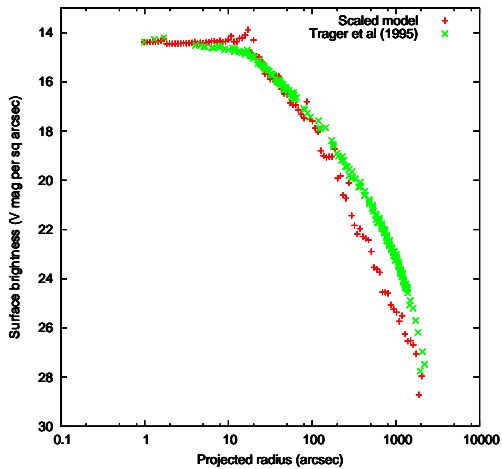}}
    \caption{Surface brightness profile of a scaled model compared
    with the composite profile of \citet{Tr1995}.}
\label{fig:sbp-scale}
  \end{figure}

\begin{figure}
{\includegraphics[height=12cm,angle=0,width=9cm]{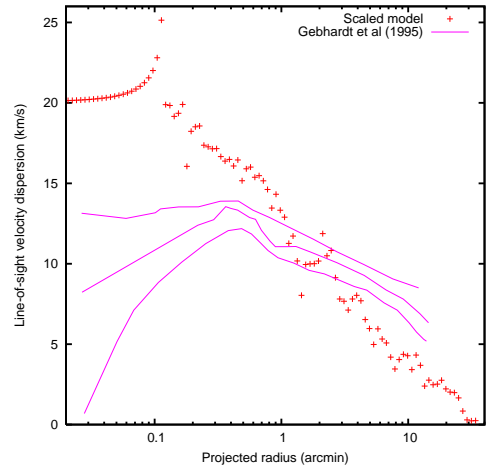}}
    \caption{Line-of-sight velocity dispersion profile of a scaled model compared
    with the observational results of  \citet{Ge1995}.  The spike in
    the model is attributable to a single star at small radius.  Only
   bright stars, with $V < 5$, were included.}
\label{fig:vdp-scale}
  \end{figure}

Figs.\ref{fig:sbp-scale} and \ref{fig:vdp-scale} compare the entire
surface brightness and velocity dispersion profiles for the model with
initial concentration $W_0 = 9$ and scaled initial half-mass radius
2pc.  It has been scaled to a system with $N = 2\times10^6$ stars. 
While the central surface brightness and velocity dispersion at
1 arcmin match reasonably well, as required, the remainder of the
profiles do not match 47 Tuc at all.  The surface brightness of the
halo is too low, and the same result is obtained if one checks star
counts \citep{DC1982}.  The model does, however, have nearly the correct core
radius.  The poor match of the velocity dispersion profile is even more
revealing.  The deficiency at large radii may be attributable to the
undermassive halo, as shown by the surface brightness profile.  The
rise in the velocity dispersion well inside the core radius, however,
is in complete contrast with the observational result.  { This}
reveals the presence in the model of a segregated population of
massive objects.  

As an aside, we consider briefly the nature of this population.  Because of the
scaling, the model has a low escape velocity, though the velocity
dispersion of natal kicks (of black holes and neutron stars) was unscaled, and so, not surprisingly, the
population of neutron stars is very small.  The compact central
population consists almost entirely of white dwarfs, which make up
49\% of the entire mass of the cluster.  There is nothing abnormal
about this, especially given the relatively flat low-mass initial mass function,
which substantially decreases the mass of the main sequence below
turnoff.  In our favoured model, however (see Table
\ref{tab:modela}), the white dwarf mass fraction is only 34\%, and the
central velocity dispersion agrees better (Fig.\ref{fig:vdp-a}).

Returning to the problem of finding a suitable scaled model, we can
also 
view the results of this mini-survey from a different angle; and it turns
out to be a more fruitful approach.  
Instead of scaling to match the
central surface brightness, { consider the result if instead }one scales to match the central
line-of-sight velocity dispersion.  { Then} it is found that the resulting
models are too { faint} centrally, and the { (observational)} core radius is too large, no
matter how compact the initial conditions are.  A better fit could
therefore be obtained if expansion of the core could be suppressed.  
Various mechanisms are known to
cause expansion of the core, including loss of mass by stellar
evolution, binary heating, and the action of black holes.  (The papers
by \citet{CW1990}, \citet{HTH2006}, and { \citet{MP2004}}, respectively, may
serve as samples of the extensive literature on these processes.)  We
judge that the initial binary fraction adopted here (2\%) is too low
{ (for our choice of initial period and mass distributions)}
to cause significant core evolution; in any case the observational
constraints (Sec.\ref{sec:pbs}) { do not suggest that it can be much
  lower}, and so we have not attempted to
reduce the binary fraction.  The remaining mechanisms for core evolution, however,
all involve the more massive stars in the initial mass function, and
we have considered modifications here.  

Despite its implausibility, perhaps, our efforts initially focused on truncation of the
high-mass IMF, by severely reducing $m_2$ ({ Table \ref{tab:ic}}).
Indeed values around 2$\msun$ gave rather satisfactory scaled models,
without any change in the slope of the upper mass function from the
canonical value of 2.3.   One problem with such a mass function is that no neutron
stars are formed
, and so we explored different ways of altering the
IMF.  Though any significant increase in the break mass always gave
poorer results, results were almost as good if we raised the maximum
mass again but compensated by  steepening the power
law index of the upper mass function.  

With only small changes in the
remaining parameters, the parameter values arrived at in this way led
to the { small-scale} model specified in Table
\ref{tab:smallscale}.  { Then}, with further changes and much
experimentation, { we arrived at}  the { full}-scale
model described in the next section.  { We summarise in Table \ref{tab:ic} the entire
ranges of all parameters which we explored, both with small-scale
models (i.e. as described in this section) and full-scale models (as in
Sec.\ref{sec:modela}).  We do not intend to suggest that all possible
joint variations in these parameters were investigated; very often the
extreme values in these ranges were tried only for one or very few
choices of the remaining parameters.  In principle it would be
desirable to adopt a more systematic way of searching this rather
large parameter space.  \citet{Ha2009} describe an  $N$-body
study of the Arches cluster which illustrates such an approach, at
least when adjusting a pair of parameters (in their case the initial
virial radius and particle number.)}


\begin{table}
  \begin{center}
\caption{An approximate small-scale model}    
\begin{tabular}{ll}
Parameter&Initial value \\
\hline
Number of stars&$N^\ast = 40000$ ($N = 2\times 10^6$)\\
 Total  mass ($\msun$) & $2.93\times10^4$ ($1.47\times10^6$)\\
 Tidal radius (pc) & 295 (109) \\
 Half-mass radius (pc) & 6.6 (2.4)\\
Central potential of King model ($W_0$)&8\\
Binary fraction & 0.02\\
Upper mass function index & 3.3\\
 Lower mass function index& 0.5\\
Minimum, break and upper mass ($\msun$) & 0.1, 0.9, 50.0\\
\end{tabular}.\label{tab:smallscale}
  \end{center}
Note: corresponding values for a full-scale model, where these are different, are given in
brackets
\end{table}

\section{A model of 47 Tuc}\label{sec:modela}


Unlike the models discussed in the previous section, the model we are
about to present is a full-scale model of 47 Tuc.   It will be seen
that the profiles and other data are considerably smoother than for
the models discussed hitherto.   This standard model will be referred to as Model A.  Details of its
initial specification, along with a summary of conditions at 12Gyr,
are given
in Table \ref{tab:modela}.  Though the slope of the upper mass
function may seem steep, still higher values (in the range 3.75--4.5) were inferred by
\citet{Me1989}.  { His result was based on the need to produce} the appropriate mass of
white dwarfs to account for the velocity dispersion profile.  Our
motivation is rather different, as we have already mentioned at the
end of the last section.  { Meylan, incidentally, took a much more nearly canonical value for the
power law index of the lower mass function than ours
(he took 1.2), but a similar break mass (he took $0.88\msun$)}.

\begin{table}
\begin{center}
\caption{Initial and present-day conditions for model A}
\begin{tabular}{lll}
&$t = 0$&$t = 12$Gyr\\
\hline
$N^1$&$2.00\times10^6$ &$1.85\times10^6$  \\
Mass ($\msun$)& $1.64\times10^6$ &$0.90\times10^6$\\
Tidal radius (pc)& 86 & 70 \\
Half-mass radius (pc) & 1.91&4.96\\
Half-mass relaxation time (Gyr)&$0.70$&$3.66$\\
$W_0$&7.5&-- \\
Observational core radius (pc)&0.42&0.55\\
Binary fraction & 0.0220 &0.0084\\
Index of upper IMF & 2.8 \\
Index of lower (I)MF & 0.40 &$0.36^2$\\
Minimum mass ($\msun$) & 0.08\\
Break mass ($\msun$) & 0.8\\
Maximum mass ($\msun$) & 50.0\\
White dwarfs ($\msun$)&0 &$3.1\times10^5$ \\
Neutron stars ($\msun$)&0 &$313$ 
\end{tabular}
\label{tab:modela}
\end{center}
Notes: $^1$ number of single stars plus number of binaries; $^2$
measured best fit between 0.3 and 0.8$\msun$, main sequence stars only.
  \end{table}

It may be of interest to note that each full-sized model such as
this takes less than a  day on a single Dual-Core AMD Opteron 2214
at {}{{2.2}}GHz.

\subsection{Surface brightness and density profiles}\label{sec:sbp}

\begin{figure}
{\includegraphics[height=12cm,angle=0,width=9cm]{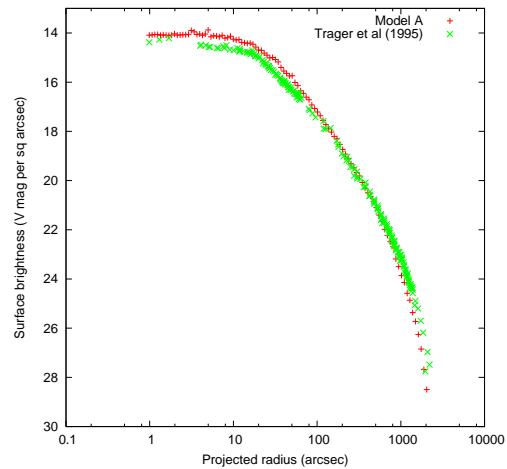}}
    \caption{Surface brightness profile of Model A compared
    with the composite profile of \citet{Tr1995}.}
\label{fig:sbp-a}
  \end{figure}

\begin{figure}
{\includegraphics[height=12cm,angle=0,width=9cm]{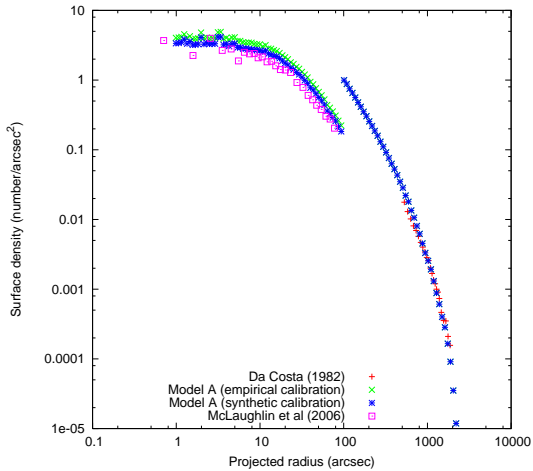}}
    \caption{Surface density profile of Model A compared with some of
    the star counts of \citet{DC1982}, which have a limiting apparent
    magnitude of $m_V = 19.45$, { and the surface density of stars
    above turnoff from \citet{Mc2006}.  Details of the ``calibration''
    are given in Secs.\ref{sec:sbp-observational} and \ref{sec:sbp}.}
    For the model, we switch from one limiting magnitude to the other
    at $100''$.}
\label{fig:sdp-a}
  \end{figure}

{ As can be seen in Fig.\ref{fig:sbp-a}, { model  A is somewhat bright at
small radii.}  Quantitatively, our central value is about 14.1 in the
units of the figure, while fits by \citet{Tr1995} and \citet{Mc2006} give values of
14.42 and 14.26 respectively.  Less evident, perhaps, because of the
steepness of the profile at large radii, is the fact that the model is
too faint there.  The mismatch approaches about 1 magnitude at the
largest radii plotted, {}{{(though this is less severe than for the scaled survey model shown in Fig.\ref{fig:sbp-scale}) and can be at least partially attributed to our treatment of the tide (see text below).}}

Star counts offer an alternative approach to the comparison of the
spatial structure of the model with 47 Tuc 
(Fig.\ref{fig:sdp-a}).  The model is somewhat overdense at small radii.  If the
cutoff in $V$ of the star counts is determined from synthetic model
data, as described at the end of Sec.\ref{sec:sbp-observational}, the
mismatch is approximately 0.1 dex.  { This} is { approximately
compatible} 
with the excess
brightness discussed above.

At large density, however, the evidence from surface density gives a
different impression from the surface brightness, as the fit with the
observational data is now much better.}
  Though the fit is
still imperfect, it was evidence
like this which for us confirmed the possibility that the composite
surface brightness profiles might { not be very reliable} at large radii.  We
had been alerted to this by finding models which produced satisfactory
mass functions (to judge by luminosity functions below the luminosity
of the turn-off) but which seemed faint at large radii when judged by
the surface brightness profile. 

 {Our treatment of the tide
\citep{GHH2008} is a further complication, as it implies that the most
distant star actually lies some distance inside the tidal radius.  For 
 this reason, the fact that {{}}} the surface  density
distribution of the model  still falls below the observations at large
radii is not necessarily a significant issue.


\subsection{Velocity dispersion profile}\label{sec:vdp}

\begin{figure}
{\includegraphics[height=12cm,angle=0,width=9cm]{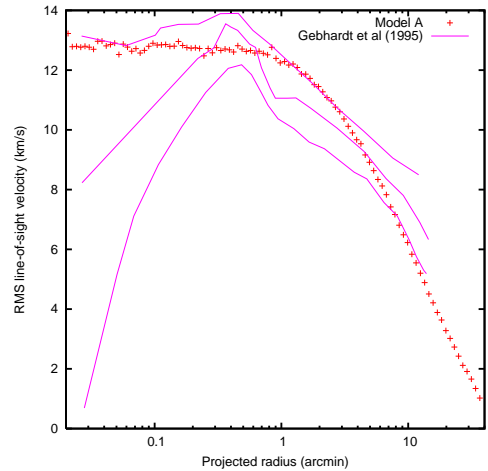}}
    \caption{Line-of-sight velocity dispersion  profile of Model A compared
    with the 90\% confidence band estimated by \citet{Ge1995}.}
\label{fig:vdp-a}
  \end{figure}

This, presented in Fig.\ref{fig:vdp-a}, is perhaps the most
unsatisfactory comparison for this model.  Whatever the formal
statistical result (by comparison with the 90\% confidence band
estimated by \citet{Ge1995}), and however much it improves on the
model shown in Fig.\ref{fig:vdp-scale}, it still gives the {
  qualitative} impression
of not conforming to the shape implied by the observations.  { This
is particularly noticeable in three places:  at the smallest radii,
where it might seem too high, and around $2'$ and  $10'$, where the
model velocity dispersion seems too high and too low, respectively.}  Besides possible shortcomings of
the model, a number of relevant factors  should be borne in mind, and
we consider them in turn.

{ Looking first at small radii, up to about $0.5'$, we see that
  {\sl any} well-fitting
  model with a relatively flat (or, at least, non-increasing) velocity
  dispersion profile must pass close to the 90\% confidence limits at
  very small radii and/or around $0.4'$. { Therefore the fact that our
  Model A does so does not cast doubt on its correctness.}

The second point to note is that, as
{ mentioned}} in Sec.\ref{sec:observed-vdp}, { our model}  takes
{ no}  account { of} the rotation of the cluster.  The maximum mean
line-of-sight { (rotational)} velocity, about 6.5km/s, occurs at a
radius of about 5--6 arcmin \citep{MM1986}; the Fabry-Perot
observations of \citet[][their Fig.11]{Ge1995} extend out to slightly
smaller radii, but they give a value above 7.5km/s already at about 4
arcmin.  Thus the rotation is large in { the second of} the three regions
where the mismatch seems most pronounced.  The influence of rotation
may also extend to larger radii:  {}{{it is worth noting that the
 single mass, rotating     Fokker-Planck cluster models of
 \citet[][their Figs 1, 11 and 12]
{Kim2008} clearly show that rapidly rotating clusters evolve faster
and have larger half-mass radii than non rotating clusters.  
This will put more mass into the outer parts of a cluster,
and increase the velocity dispersion there.}}

The third factor to
bear in mind is the influence of the tide at large radii.  We treat
its action as a cutoff, which is an increasingly rough approximation
as the tidal boundary is approached.  (The observations of
\citet{Ge1995} extend to almost half the observed tidal radius.)  It
has been known since the work of \citet{Dr1998} that the velocity
dispersion profile in some clusters flattens off towards the 
{}{{tidal}}
radius, presumably because of the effects of the galactic tide.  Such
an effect is certainly observable in $N$-body simulations
{}{{\citep[e.g.][]{GH1997,CD2005,KKBH2010}}} 
and has also been confirmed recently 
 for 47 Tuc by \citet{LKLISBS2010} (see Fig.\ref{fig:rv}). Therefore the apparently excessively
rapid decline of the velocity dispersion at large radii in the model may be an
artefact of our simple tidal treatment.  We note in this respect that
the multi-mass anisotropic King model of \citet{Me1988} did a much
better job in fitting the outer velocity dispersion and its trend.  The very
large half-mass radius in that model {(Sec.\ref{sec:size})} is no doubt a factor, placing
more mass at large radii and elevating the velocity dispersion there.
The fact is, however, that our models, which start as compact King models, do
not evolve as far as  those constructed by Meylan.  { Also, it
  seems likely that some of the velocity dispersion which Meylan's
  model fitted is due to processes which were not included in that
  model, i.e. the tidal field.}

 It is worth reporting here that we have attempted, with very limited
success, to improve the outer velocity dispersion by constructing
models which might enhance the mass density at large radii.  In particular
we tried 
\begin{enumerate}
\item  Woolley models
\citep{Wo1954}, which  employ a truncated isothermal distribution, in
contrast to the lowered isothermal distribution of the King model;
their concentration 
can be specified in much the same way, i.e. by the scaled central
potential $W_0$; 
\item polytropic models, with polytropic index exceeding the Plummer
  value of 5; the models were truncated at the tidal radius and
  revirialised globally, just as would be done with a Plummer model; and
\item 
{ initially mass segregated models generated according to either (a)
the prescription given by \citet{MV2007}, with time delay of stellar
evolution up to 250 Myr; or (b) the method of \citet{BDK2008}, with mass segregation
down to 2 $M_{\odot}$.}
\end{enumerate}
{Mass segregated models do differ in the central part of the system,
generally slightly lowering the velocity dispersion, but in the region
farther out than about $5'$ they are indistinguishable from { models
  without primordial mass segregation. }}
{ Polytropic models in particular provided a slightly better fit to the
velocity dispersion profile, { though the improvement is too
  marginal to justify abandoning King initial conditions.} } 

  
\subsection{Luminosity  and mass functions}\label{sec:lf-a}

\begin{figure}
{\includegraphics[height=12cm,angle=0,width=9cm]{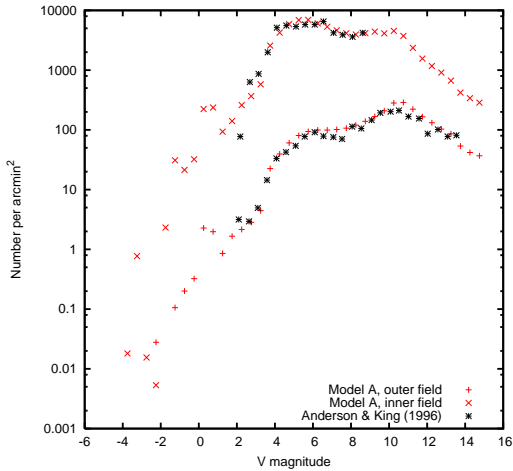}}
    \caption{Luminosity functions at two radii, in Model A and from
    the observational data of \citet{AK1996}.  Note that these authors
    specify the radii in terms of the core radius (1 and 14$r_c$), and
    we have assumed $r_c = 23''$.}
\label{fig:lf-a}
  \end{figure}

\begin{figure}
{\includegraphics[height=12cm,angle=0,width=9cm]{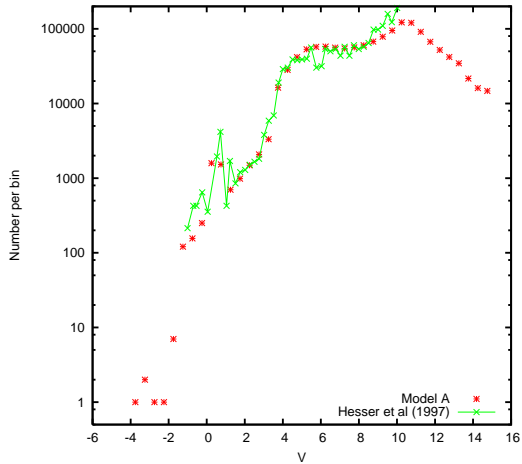}}
    \caption{{The global luminosity function of model A in comparison
    with that of \citet{He1987}.  The latter has been normalised to
    give a reasonable fit in the upper main sequence.}}
\label{fig:globallf}
  \end{figure}

Luminosity functions in fields at two radii are shown in
Fig.\ref{fig:lf-a}, in comparison with data from \citet{AK1996}.
According to those authors, turnoff corresponds to about $V = 4.1$ on
the scale of this diagram.  Below turnoff, then, the agreement is
fairly satisfactory in both fields, in terms of both overall shape and 
normalisation.   {The quality of the normalisation is relevant to the
discussion in Sec.\ref{sec:sbp}.  For instance the inner field is at a
radius of about $23''$ where, as we have seen, the evidence of both
number counts (for stars above turnoff) and surface brightness is that
the model is too dense or bright by about 0.1 dex.  Nevertheless the
mean difference in the logarithm of the luminosity function in
Fig.\ref{fig:lf-a} between $V = 4$ and 8 is $0.02\pm0.03$ (standard
error), i.e. the model is not significantly denser than the
observations. 

This discussion is complicated by the fact that the data of
\citet{AK1996} do not extend much above turnoff, whereas the surface
density data discussed in Sec.\ref{sec:sbp} is confined to stars above
turnoff, and they also dominate the surface brightness.  To explore
the luminosity function above turnoff, in Fig.\ref{fig:globallf} we
compare our global luminosity function with the composite luminosity
function of \citet{He1987}.  We have somewhat arbitrarily normalised
the latter (by eye) in the upper main sequence, where we know from the
evidence of Fig.\ref{fig:lf-a} that the model appears to be
satisfactory.  A satisfactory match continues up to about the
prominent peak (at the location of the horizontal branch), but no
further.  Actually, the distribution
along the giant branch is known to be a sensitive test of stellar
evolution \citep{BV2001}, complicated in 47 Tuc by the spatial
variation of the giant branch itself \citep{Fr1985,Ba1994}.  Though we
therefore make no attempt to resolve the mismatches in this brightest
part of the luminosity function, it plays a significant role in the
surface brightness, and problems here may well contribute to the fact that
the surface brightness and density of the model appear to be excessive.}


Let us return again to 
 a consideration of stars below turnoff.  
{ The low-mass mass function is often discussed in terms of the
power-law index over some convenient mass range.   We have preferred to carry out the
comparison in the observational domain, i.e. in terms of the
luminosity function, partly because the mass range over which the
power law is fitted varies from one author to another, and partly
because it is not always clear how the fit is actually carried out.  In any
event, such comparisons depend on the choice of mass-luminosity
relationship.  Nevertheless we give one example, which is a comparison
with the ground-based result of \citet{He1987} for the global mass
function.  
They found an index
of $1.2\pm0.3$ (Salpeter = 2.35) for masses corresponding to
 $5\ltorder V\ltorder 9$, which corresponds to masses between about 0.5 and 0.8$\msun$.  We find at least
that an index of $0.9$ provides a reasonable fit in this range.
{  Nevertheless, a smaller value provides a better fit (in our model)
 if the range is extended to include more of the lower main sequence
 (Table \ref{tab:modela}).}  Indeed the change from the primordial
 value is less than 0.1.  \citet{BM2003} used $N$-body models (not
 specifically geared to 47 Tuc) to
 assess the dependence of the mass-function index on mass lost by the
 cluster; and our result is entirely consistent with their Fig.9,
 since in our model 47 Tuc has lost less than half of its initial mass.}




\subsection{Pulsar accelerations}

\begin{figure}
{\includegraphics[height=12cm,angle=0,width=9cm]{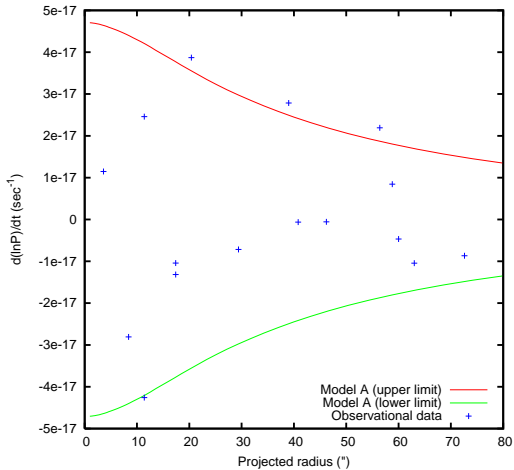}}
    \caption{Comparison between spindown rates of millisecond pulsars \citep{Fr2003}
    and the extreme contributions from the gravitational
    acceleration of Model A.  }
\label{fig:pulsars-a}
  \end{figure}

Fig.\ref{fig:pulsars-a} shows that the model is very nearly
consistent with observations of the spin-down of
pulsars in 47 Tuc.  
  The fact that three
lie just above the upper curve is consistent with estimates of the
small intrinsic component (Sec.\ref{sec:pulsars}). { The critical
object in this plot is the pulsar with the largest negative period
derivative.  It is known as 47 Tuc S \citep{Fr2003}, and these authors
showed that it implies a projected mass/light ratio $M/L > 1.4$ in the
region of the pulsar.  In fact the projected value of $M/L$ at the
location of this pulsar in our model is about 1.1.  Note, however,
that our model is a little bright in the core (Sec.\ref{sec:sbp}),
which depresses the value.  (Incidentally, the projected value of
$M/L$ increases from this central value to about 2.3 at large radii;
the global value is about 1.52.)
Generally speaking, the tension between the requirements of the
central surface brightness, on the one hand, and the acceleration of
47 Tuc S, on the other, was the single most important constraint in
our  attempts to find a satisfactory model.  
}


\subsection{Variation of Parameters}\label{sec:variations}

{{ Even though we had to work very hard to reach Model A, and discarded
  numerous models in the process,  we can make no  claim} to the uniqueness of the model presented
above, even within the limitations of the parameters which
characterise our initial conditions.  But in this subsection we wish
to summarise the most notable changes { (in the surface brightness,
  surface density, velocity dispersion profile, mass functions and
  pulsar accelerations)} which occur if we adjust our
choice slightly.  The list of parameters begins with those in Table
\ref{tab:modela}, but we have also considered a number of other
possibilities, which are included in the list below.  
Where we have varied a given parameter, the values of the remaining
parameters are fixed and close to, but not always exactly the same as,
those in Model A.  Where no mention is made of changes, it means that the changes (in
  the five quantities listed above) are
negligible in comparison with other entries.  We generally desist
from giving a theoretical interpretation of these empirical
statements, each one of which might be a research project on its own.

\begin{enumerate}
\item $N$ (Initial number of stars):
an increase by 5\% leads to an increase in the velocity dispersion by
about 3\%, and an increase in the maximum gravitational pulsar spinup
by about 4\%.
\item $r_t$ (Initial tidal radius):
an increase by about 25\% leads to a decrease in the central surface
brightness by about 0.3 mag and a corresponding decrease in the
luminosity function in the inner field, an increase in the outermost radius in
the surface brightness and density profiles by about 10\%, a decrease
in the central line-of-sight velocity dispersion by about 5\%, and a
decrease in the maximum gravitational pulsar spinup by about 25\%.
Note that, because we hold the initial ratio of tidal and half-mass
ratio constant, a variation of the initial value of $r_t$ also affects
that of $r_h$.
\item $r_t/r_h$ (Initial ratio of tidal and half-mass radii):
An increase by 12.5\% (i.e. a correspondingly more compact initial
configuration, relative to the tidal radius) gives an increase in the
central surface brightness by about 0.1 mag { per square arcsec}, an increase in the
central velocity dispersion by about 4\%, and an increase in the maximum
gravitational pulsar spinup by about 18\%.
 \item $W_0$ (Initial King concentration):
an increase by 1 leads to an increase in the maximum gravitational
pulsar spinup by about 14\%, but no other significant changes.  
This is also discussed in
 Sec.\ref{sec:finding-a-model}; see especially Fig.\ref{fig:survey}.
\item $f_b$ (Initial binary fraction):
an increase by 50\% (e.g. from 0.02 to 0.03) leads to a 
decrease in the central surface brightness by about 0.3 mag, a
decrease in the central surface density by about 19\%, and a decrease
in the maximum gravitational pulsar spinup by about 11\%.
\item $\alpha_1$ (Index of the lower initial mass function):  
an increase by 0.1 (e.g. from 0.4 to 0.5) decreases the velocity
dispersion profile by about 2\%, and causes small changes in the outer luminosity
function as expected from the discussion in
Sec.\ref{sec:finding-a-model} (see especially Fig.\ref{fig:imf-comparison}).
\item $\alpha_2$ (Index of the upper IMF): 
an increase by 0.2 (e.g. from 2.8 to 3.0) leads to an increase in the
central surface brightness by about 0.4 mag, an increase in the
central surface density by about 0.15 dex, and an increase in the 
maximum gravitational pulsar spinup by about  18\%.
\item $m_b$ (Break mass):
an increase from 0.75 to 0.85$\msun$ leads to a decrease in the
central surface brightness by about 0.6 mag, a decrease in the central
surface density by about 30\%, some changes (in line with naive
expectations from the change in break mass and the smaller central
surface density) in the inner luminosity function, and a decrease in
the maximum gravitational pulsar spinup by about 20\%.
\item $m_1$ (Minimum initial mass):
we did not experiment widely with this parameter, but can report that
an increase from 0.08 to 0.1$\msun$ causes an increase in the core
velocity dispersion by about 2\%.
\item $m_2$ (Maximum initial mass):
a change from 50 to 150$\msun$ leads to negligible changes.
\item $t$ (Time [Age]):
an increase by 1Gyr leads to a decrease in the central surface
brightness by about 0.15 mag, a decrease in the central surface
density by about 20\%, and a decrease in the velocity dispersion by about 6\%.
\item $D$ (Distance):
an increase by 0.5kpc leads to a corresponding reduction in the
apparent length scale of the surface brightness, surface density and
velocity dispersion profiles, and an increase in the inner luminosity
function by about 8\%.
\item $Z$ (Abundance):
an increase by 0.0005 from 0.003 to 0.0035 leads only to a decrease in
the maximum gravitational pulsar spinup by about 4\%.
\item $\sigma_k$ (One-dimensional dispersion of natal kicks of neutron stars):
a decrease from 190km/s (our canonical value) to 160 km/s leads to no
significant changes, {}{{but  a larger decrease to 130 km/s
    leads to an increase in the maximum gravitational pulsar spinup by about 10\%}}.
\end{enumerate}
}


\section{Discussion of the model}

\subsection{The size of 47 Tuc}\label{sec:size}

{ A description  of the spatial structure of globular clusters is often
  reduced to just three values: the core, half-mass and tidal radii.}   Our value for
the tidal radius (Table \ref{tab:modela}) is rather comparable with  those of
\citet[][about 70 pc]{Me1988} and Table \ref{tab:data} (which converts
to 56pc at the stated distance).    Our value for the observational core
  radius (Table \ref{tab:modela}) is quite similar to that derived
  from Table
  \ref{tab:data}, i.e. 0.52pc. 
There is, however, an amazing variation in quoted values for the half-mass
radius.  Data in Table \ref{tab:data} leads to a value of about 3.65
pc, and the difference from our value (Table \ref{tab:modela}) may be
explicable if the former value should really be understood as a
projected half-light radius.  The values in Meylan's best models,
however, lie close to 9.5pc, and we have no explanation for such a
wide disagreement.  It does, however, have the effect of placing more
  mass at larger radii in his model, and this may be one reason why
  his fit with the velocity dispersion profile at large radii is more
  successful than ours (see Sec.\ref{sec:vdp}.)

\subsection{Velocity anisotropy}\label{sec:va}

\begin{figure}
 {\includegraphics[height=12cm,angle=0,width=9cm]{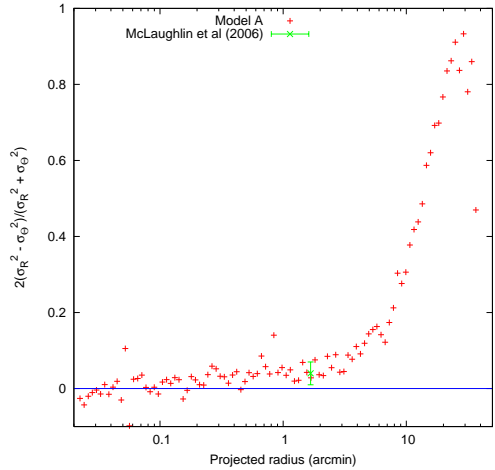}}
     \caption{Anisotropy in the proper motions of Model A and the
       observational 
       result for stars with $m_V < 18.5$ within $100''$ \citep{Mc2006}.
 The label on the ordinate specifies what is computed, the mean square
{ radial and tangential}       proper motions at a particular line of sight being $\sigma_R^2$
       and $\sigma_\Theta^2$.   }
 \label{fig:anisotropy}
   \end{figure}

The velocity anisotropy of 47 Tuc near the centre has been { constrained
observationally }
by \citet{KA2001} and \citet{Mc2006}.  The former authors also
measured anisotropy at about $4'$ from the cluster centre, but gave no
quantitative results.  Fig.\ref{fig:anisotropy} shows anisotropy among
bright stars ($V < 6$) in our model, compared with the result for the
brightest stars ($m_V < 18.5$) within about $100''$ of the centre
\citep{Mc2006}.  Though the observational result differs only
marginally from zero, the results are consistent.  A rapid rise at
large radii, as seen in our model, is qualitatively consistent with
what was inferred indirectly by \citet{Me1988} on the basis of model fitting.

\subsection{The binary frequency}

\begin{figure}
 {\includegraphics[height=12cm,angle=0,width=9cm]{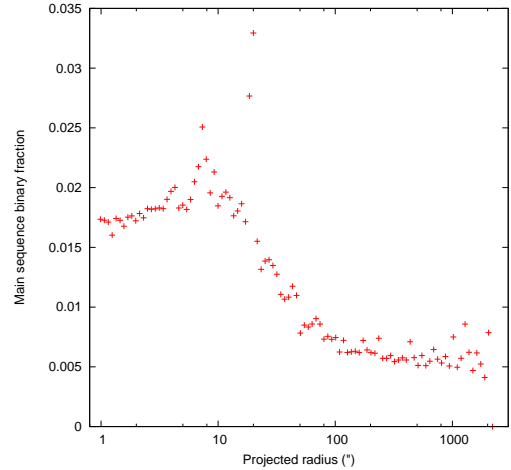}}
     \caption{Projected fraction of main sequence binaries (as defined
       in the text) { in Model A}. {The high points are an
       artefact of the way in which the surface density is computed,
       which can be very large if the line-of-sight nearly coincides
       with the radius of a binary.}}
 \label{fig:fb}
   \end{figure}

Fig.\ref{fig:fb} shows the radial distribution of the fraction of main
sequence binaries.   { This was calculated as in Sec.\ref{sec:sdpbflf}, except
  that in calculating the surface density of single stars we include
  only main sequence stars, and in the surface density of binary stars
  we include only  binaries in which both  components are  main
sequence stars.  In both cases also the apparent $V$ magnitude was
limited to the range 20.5 to 23.5.  These are quite appropriate
choices  for comparison with the observational data summarised in
Sec.\ref{sec:pbs}, except that the range of magnitudes used by
\citet{Mi2008} was a 3-magnitude range specified in $I$.  In the
radial range of their data, our result is just consistent with theirs.}

\begin{figure}
 {\includegraphics[height=12cm,angle=0,width=9cm]{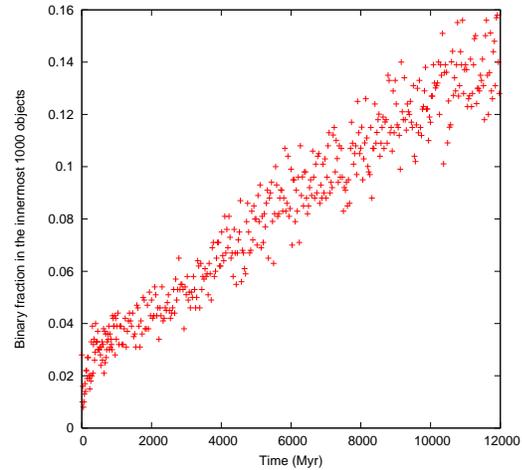}}
     \caption{ Fraction of all binaries in the innermost 1000 objects
     { in Model A}.}
 \label{fig:fb-1000}
   \end{figure}

Of greater relevance for dynamics is the 
fraction of all binaries, not just those with main-sequence
components.  { Our initial conditions generate a large proportion of
relatively soft pairs.  These are quickly destroyed, and the global
fraction decreases by over 60\% (Table \ref{tab:modela}).  Within} the innermost 1000 stars,
however, the binary fraction increases with time as
recorded in Fig.\ref{fig:fb-1000}, { because of mass segregation.}

\subsection{The dynamical role of dark remnants}

In the core, the velocity dispersion in the model is certainly
affected by degenerate components, which make up about 51\% of the
mass there, but their mass fraction decreases at larger radii to
give a global ratio of about 34\% (see Table \ref{tab:modela}).  At face value, our result for 
the core 
contrasts with the finding of \citet{Mc2006}, who suggested, on the
basis of the { profile of dispersion of proper motions}, that the contribution
of heavy remnants (such as neutron stars and white dwarfs) is only a
fraction of a percent.  The resolution of this discrepancy may simply
be to understand that their estimate applied only to neutron stars,
and perhaps the most massive white dwarfs.  Our global result is
comparable to a number of earlier  estimates, based on model fitting,
which were summarised by \citet{HH1996}.

{ 
{ Despite our tinkering with the initial mass function, and despite
  giving natal
kicks with a one-dimensional dispersion of 190km/s, our model
appears to produce a reasonable number (213) of neutron stars, .}  A recent indirect
estimate, based on a quite simple demographic argument, puts the total
neutron star population in the cluster at ``$\sim$300''
(\citet{He2005}; { see also \citet{VM1988}}).   It has
been argued \citep{Iv2008} that substantial numbers of neutron stars must be formed
in processes leading to smaller natal kicks than assumed in our model,
and so the number in the model could well be larger if this refinement
were added.  

}

In our model there are only 19 stellar-mass black holes at the
present day, and their total mass is small compared with that of other
remnants.  The maximum number present at early times is over 1000, but
the number falls to less than 40 within the first 30Myr. 

{There is no sign of any early phase of runaway coallescence in
  the model.}  { The results of \citet[their eq.(17)]{PZ2002}, based on theory
  calibrated by $N$-body models up to $N = 64k$, imply that the
  collision rate is less than one per Myr.  They also noted, however, that their result
  is likely to be an overestimate at half-mass relaxation times above
  about 30Myr.  Furthermore our upper mass function is steeper than
  theirs, and this helps to diminish the rate of collisions.
  Incidentally, the initial half-mass density, which is
  $0.28\times10^5$, is within the range of a number of young massive
  clusters at the present day \citep{PZ2010}.
}






\subsection{The dynamical status of 47 Tuc}\label{sec:status}

Because of its high concentration, it is tempting to assume that 47 Tuc is
close to core collapse.  Though it is a massive cluster, its core
relaxation time is given \citep{Ha1996} as  $t_{rc}\simeq9\times10^7$yr.  For
an equal-mass King model having the same concentration as 47 Tuc at
the present day, i.e. about 2.03, the core collapse time is about
$400t_{rc}$ \citep[from data in][]{Qu1996}.  But it is known
\citep[e.g.][]{CW1990} that the time to core collapse in systems with
unequal mass is much smaller than that in equal-mass systems; these
authors quote an example of a $W_0 = 7$ King model, where a multi-mass
system collapses in a time smaller than an equal-mass system by a
factor of approximately 45.  Putting all these factors together
suggests that, indeed, the core of 47 Tuc should collapse within about
1Gyr.   The large numerical factors involved, however, suggest that
this argument is quite precarious.

   \begin{figure}
 {\includegraphics[height=12cm,angle=0,width=9cm]{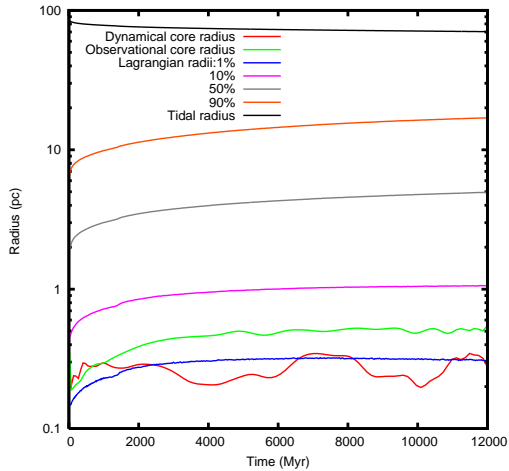}}
     \caption{ Evolution of lagrangian, tidal and core radii in Model
     A.  { Definitions of the core radii are given in the text.}}
 \label{fig:radii}
   \end{figure}

\begin{figure}
 {\includegraphics[height=12cm,angle=0,width=9cm]{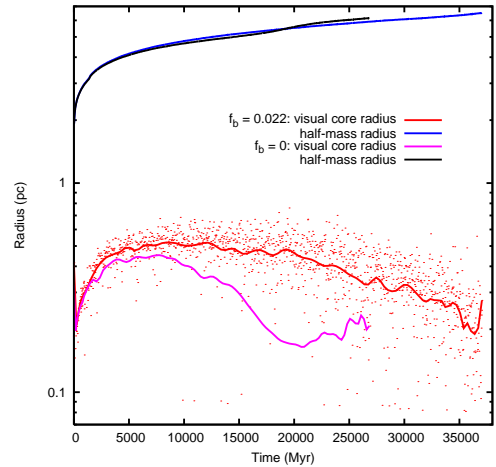}}
     \caption{ Evolution of half-mass and observational core radii in
     Model A and in a model with no primordial binaries but which is
     otherwise identical.  The core
     radii are smoothed, but the original model data are also shown
     for Model A.  Both runs were stopped at the point where  the data
     ends.}
 \label{fig:comparison}
   \end{figure}

The spatial evolution of Model A is shown in Fig.\ref{fig:radii}.  The
core radius is represented by the two curves, one of them quite wavy,  near the bottom
of the figure, not far from the 1\% Lagrangian radius.  These two curves are
{ Bezier}  fits to rather noisy data on the core radius {
  (cf. Fig.\ref{fig:comparison})} generated
in two ways.  The curve labelled in the key as ``dynamical'' is
determined from the mass-weighted three dimensional  velocity
dispersion and  density of the innermost
20 particles, using the formula 
$
r_c^2 = {3\langle v^2\rangle}/({4\pi G\rho_0}).$  The ``observational''
curve is determined from a computation of the { surface brightness
profile}, and is the radius at which this falls to one half of its
central value { (Sec.\ref{sec:sb})}.

We first notice that, except at the very start, there is a 
fairly steady (though decelerating) increase in the observational
core radius, at least in absolute terms.   
It also increases relative to the dynamical core 
radius within the first few Gyr.  
We interpret {  the increase relative to the
dynamical core radius} as due to a combination of stellar
evolution and mass segregation: as the turnoff mass decreases, the
observable core is determined by stars of lower mass, which are more
spatially dispersed.  \citet{Hu2007} has described a similar effect in
$N$-body simulations, but attributed it to the core collapse of the
heavy (and presumably dark) stars.  Perhaps both effects are present,
or perhaps they are simply somewhat different descriptions of the same
processes.

{ After the first few hundred Myr, the dynamical core radius is almost constant in absolute terms, but
does appear to decrease slowly with respect to the half-mass radius
(Fig.\ref{fig:radii}).  The time scale of this
process, which we refer to as core collapse, is difficult to estimate
from Fig.\ref{fig:radii}, but the predicted future evolution is given
in Fig.\ref{fig:comparison}.  Evidently the collapse of the core
slowly accelerates, but continues for at least a further 25Gyr!}
On this evidence, then, 47 Tuc
is far from core collapse.

Now we turn to the mechanism which causes the overall expansion of the
system, the fact of which is evident from Fig.\ref{fig:radii}. Three
processes come to mind:
\begin{enumerate}
\item mass loss from stellar evolution;
\item primordial binaries; and
\item new binaries, formed in three-body interactions.
\end{enumerate}
{ Empirical evidence comes from a comparison simulation in which the
initial binary fraction was 0 (Fig.\ref{fig:comparison}); we refer to
this as Model A0.  The
half-mass radius increases at almost the same rate in both models,
even though Model A0 contains at most one
binary (formed in three-body interactions) until it reaches an age of
about 23Gyr { (a few Gyr after the estimated time of core bounce, at about 20Gyr)}.   There are no binaries in it between about 500Myr and 4Gyr,
and yet the expansion of the half-mass radius of  Model A0 is very
close to that of Model A, even during this period.  It must be
concluded, therefore, that the effect of binaries is almost negligible
in both models, and almost the entire expansion must be due to mass loss from
stellar evolution,  
at least up to an age
of 12Gyr.  

The difference between the
two models is more noticeable in the core, however, and we conclude
that the core of 47 Tuc would be about 20\% smaller were it not for
the action of the primordial binary population.  In  Model A0 it seems
clear that the core radius has
passed its maximum by an age of 8Gyr, and the figure shows that core
collapse is arrested at about 20Gyr.  From then on three-body
binaries form in greater abundance, and there is a slight but
noticeable acceleration in the expansion of the half-mass radius.}



{ It may seem surprising that mass loss from stellar evolution is
  dominant.   Its time scale, however, is set by astrophysics.  The other
  possible competitors (core collapse, and heating by primordial
  binaries) scale with the relaxation time, other things being equal.
  This time scale increases with the mass of the cluster, and so for a
  sufficiently massive cluster, these stellar dynamical processes are
  relatively weak.  47 Tuc ranks fifth among the globular clusters of
  the Galaxy for luminosity \citep{Ha1996}, and, given the rather
  narrow range of $M/L$ ratios, is certainly one of the most massive.

It is an interesting and puzzling fact that the most luminous Galactic
globular clusters tend to be the most concentrated \citep{vB2003}.  It
is puzzling because, if we ignore stellar evolution, we expect the clusters of lowest mass to evolve
fastest and therefore to approach or enter core collapse sooner;  and
this is the opposite of what is found.  We see from this model of 47
Tuc, however, that high concentration is not necessarily any
indication that the cluster is close to core collapse.}

\subsection{Missing ingredients}\label{sec:missing}

Several realistic aspects of star cluster evolution are omitted from
our modelling.  These include rotation, oblateness effects, a
time-dependent tide (due to the ellipticity of the Galactic orbit of
the cluster), tidal accelerations (as opposed to a cutoff), various
aspects of binary dynamics (because we use cross sections for the
interactions), and the complexities of initial conditions.  Some of
these omissions are common to virtually all simulations of star cluster
dynamics, and have been since the field began.  At the present day,
however, especially in the context of globular star clusters, it is
the question of initial conditions which is most controversial and
most fluid.  

Initial conditions are complicated by a number of factors, of which we
highlight two.
The first is early gas expulsion.  { This} has
been the subject of considerable research for some time now 
\citep[e.g.][]{Ma2008,Ba2008,GB2006,BG2006,FK2005,GB2001}, but we have
ignored its effects.
The second problem is that it has to be recognised
that 47 Tuc cannot really be regarded as consisting of a single
stellar population any more \citep{An2009}.   The appropriate
paradigm for this situation, let alone its simulation, are full of
uncertainties \citep{DE2008,CY2008,Do2007}.  

Faced with these
complexities one might conclude that a modelling exercise like that in
the present paper is meaningless, misguided or at least premature.
Our view on this is based on the fact that these missing features may
affect only the first hundred million years of evolution, and perhaps
less.  
We make
the assumption that our models, after such a period of evolution, do
resemble hypothetical  models which start with different initial
conditions but do include these missing ingredients. { Thus}
\citet{Ma2008} have shown that models which include 
gas expulsion produce a mass function which
approximately resembles what we have chosen for our IMF (at least
qualitatively, { in the sense that the lower mass function will have an
index { smaller} than the canonical value of around 1.3}).  

 These processes are likely to depend on the mass of the cluster, and
it is not surprising that our choice of IMF for 47 Tuc is
quantitatively somewhat different from those we preferred for the
clusters NGC6397 and M4 \citep{HG2008a,GH2009}, which we consider to
have been less massive initially by a factor of { about} four.  {  But the
length scale is also important.  { It is interesting to note, for
instance, that the central density of our initial conditions for these
three clusters lies within a fairly small range (Table
\ref{tab:central}).  { The same is true of the central escape
  speed.  Indeed,} for most of its life so far, the central
escape speed in 47 Tuc is larger than for the other two clusters by
a factor of only about two \citep[see][Fig.12]{GH2009}, though it
still exceeds 40km/s at the present day.  If chemically contaminated winds are a
significant factor in the production of a second generation, the
retention factors among different clusters may not vary over a wide
range.}
 

\begin{table}
  \begin{center}
    
  \caption{Initial central conditions for models of three clusters}
  \begin{tabular}{llll}
    Cluster&Central density &Core radius&Escape speed\\
&($\msun$/pc$^3$)&(pc)&(km/s)\\
\hline
M4&$0.9\times10^6$&0.31&81\\
NGC 6397&$3.0\times10^6$&0.22&101\\
47 Tuc&$1.0\times10^6$&0.37&116
  \end{tabular}\label{tab:central}
  \end{center}
\end{table}
}

{ When this project of fitting Monte Carlo models to individual globular
clusters was begun, it was thought that post-collapse clusters would
be most difficult to treat, because of the complex interaction of
several processes which would lead to their present structure.  Our
conclusion now is that pre-collapse clusters are equally difficult,
because they have changed less from their virtually unknown initial
conditions.}

\section{Conclusions}

We have constructed a dynamical evolutionary model of the massive
Galactic globular cluster 47 Tucanae (NGC104).  The model takes into
account dynamical interactions between stars and binaries, the stellar
evolution of these components, and the effect of the Galactic tide.
{ We make no claim for the uniqueness of this model, though
numerous other models were eliminated in the search for it.}

The model begins with a moderately concentrated King model {($W_0 =
7.5$)} without
primordial mass segregation.  { The initial and
present-day characterisation of the model, including values for its
mass, spatial dimensions, the mass of degenerate remnants, and the
binary fraction, are given in Table \ref{tab:modela}.  In summary, the
initial mass, half-mass radius and number of stars are
$1.64\times10^6\msun, 1.91$pc and $2.00\times10^6$, respectively.  By
the present day these values have changed to $0.90\times10^6\msun,
4.96$pc and $1.85\times10^6$, respectively.   The most unorthodox aspect of the initial
conditions is the mass function, which is steeper than Salpeter for
high masses, and relatively flat at low masses.}

We judge the success of the model by
comparison with observations of 47 Tuc at the present day:  the
surface brightness and surface density profiles, the velocity
dispersion profile, the luminosity function at two radii, and observed
pulsar accelerations.  The centre of the model is a little bright,
{ perhaps because of some flaw in the recipes for stellar evolution
  of post-main sequence stars.}  
The least satisfactory comparison is with the
velocity dispersion profile.  On the other hand it is not clear how to assess the
velocity dispersion profile of a spherical, non-rotating model which
is constructed so as to resemble the spatial structure of a somewhat
flattened, rotating object.     

We find that
the primordial binary population of 47 Tuc is playing a significant
role in the evolution of the core radius but not, at present, a
dominant one.  The cluster is far from core collapse, which will not
take place for another 2{5} Gyr at least.

\section*{Acknowledgements} 
We are indebted to M Bagchi for several very instructive conversations
about the pulsars in 47 Tuc,  to P Hut for fruitful discussion
about the processes behind the evolution of our model,{ and to
J. Kaluzny for advice on the age and distance of 47 Tuc}.  { We are
grateful also to the referee for helping us to organise and express
our thoughts a bit better.}  DCH thanks his
host, S Mineshige, for hospitality at Kyoto University during a visit
in which this work progressed a lot.  The visit was supported by the
Grant-in-Aid for the Global COE Program ``The Next Generation of
Physics, Spun from Universality and Emergence" from the Ministry of
Education, Culture, Sports, Science and Technology (MEXT) of Japan.
{ He also warmly thanks his coauthor for his hospitality during a visit
to CAMK.} 
{MG was supported by the Polish Ministry of Sciences and Higher Education through the grants 92/N-ASTROSIM/2008/0 and N N203 38036.}

\bsp

\label{lastpage}

\end{document}